# GUI Based Automatic Remote Control of Gas Reduction System using PIC Microcontroller

**Ayad Ghany Ismaeel**
Erbil Technical college – Hawler
Polytechnic University (previous FTE-
Erbil), Iraq.
Email: dr_a_gh_i@yahoo.com

**Raghad Zuhair Yousif**
Salahaddin University – Erbil-Iraq,
Email: rag_zuh@yahoo.com

**Essa F. Abdallh**
Erbil Power Station
Email: essa_faik@yahoo.com

*Abstract—the GRS is a one of the important units in Erbil Power Station EPS, which is responsible on controlling gas pressure and gas temperature this unit previously works manually. The local control panel for GRS system contains two types of digital signals the first one indicated by Light Emitting Diodes LED to point normal operations, fault and alarm, and event of operations while the second indicated by ON-OFF switches, which consists of two types the push buttons switch and mode selector switch. To overcome human in manual control faults in controlling GRS systems, automation system becomes the best solution. The purpose of this research is to design and implement embedded automation system that can be used to control a GRS automatically through a GUI and from remote location by using programmable interface controller (PIC16F877A). In this research the (PIC) software which is based on (C language), developed by Microchip (MPLAB) is used in programming a PIC microcontroller, then Visual Basic is used in the construction of GUI, the RS-232 serial cable is used as a connector between PIC and PC. Implement the proposed design and test it as a first system shows all operations of GRS successful were converted into full computerize controlling (with the ability of full automatic control) from remote location through proposed GUI.*

*Keywords-Peripheral Interface Controller (PIC); Microcontroller; Graphical User Interface (GUI); Remote; Control*

## I. INTRODUCTION

The control and monitoring machines from remote location are very important in these days due to increasing the factories and plants [6]. The automation systems and embedded control systems are used when the accurate and quick decision required, and when the human life being in hazard for doing some jobs inside electrical power stations and chemical plants. (hazardous working environments) [1].

In this work the procedure of design and implementation embedded automation software and hardware integrated system for the tasks implemented by Gas Reduction system GRS machine are:

1) Separator System: The main function of the fuel gas filter/separator is to remove solid contaminants and liquid (condensed gas constituents) from the incoming gas flow. The filter elements are designed for continuous operation with max temperature of 60°C.
2) Boilers: provide desired heat for the water system
3) Pressure Regulator: regulate the pressure of Gas through adjusting valve heat into desired value.

4) Hot Water System: produce hot water for the gas heating system.
5) Local Control Panel: receive all status and alarm signals from the fuel gas supply system.

Remotely control of GRS as a unit(s) working in Erbil Power Station EPS is presented. The GUI resident on a control computer was built based on Visual Basic, its used to monitor and control the GRS unit from remote location which leads to limiting the probability of Fault and error in the system manual operation, moreover adding automatic operation mode as a command Push button to the machine GUI allows automatic operations without human interactions. The presented software package helps the control engineer in recognizing the same switches and indicators in the real panel easily. The proposed system integrates software and hardware tools in its architecture.

*A. Functional Description of GRS machine*

The GRS is a Gas Reduction System which is designed to receive the natural gas from the incoming gas pipelines, treat and condition the gas to meet the operating conditions specified by the manufacturer of the gas consumers. The GRS works in power station, the power station use the output of GRS as a fuel of its operation. One of the GRS system units works in EPS is shown in Fig. 1.

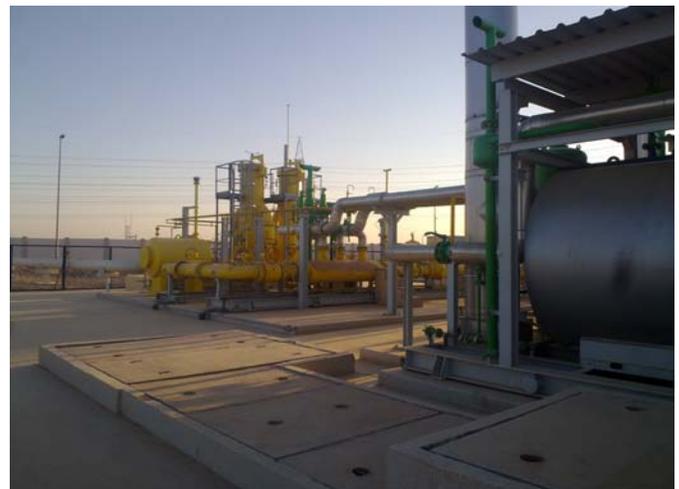

Figure 1. GRS working in EPS

*B. GRS Units and Operation*

The GRS system contains boilers for heating the inlet gas into desired set point. Fig. 2 shows the local control panel inside GRS system. The GRS control panel contains two categories of digital signals the first category of signals are indicated by LED Light Emitting Diodes while the second category of digital signals is handled by ON-OFF switches. There are three LED colors have been used .Green LEDs means normal operations, Red LEDs mean fault and alarm operations, Yellow LEDs mean event of operations.

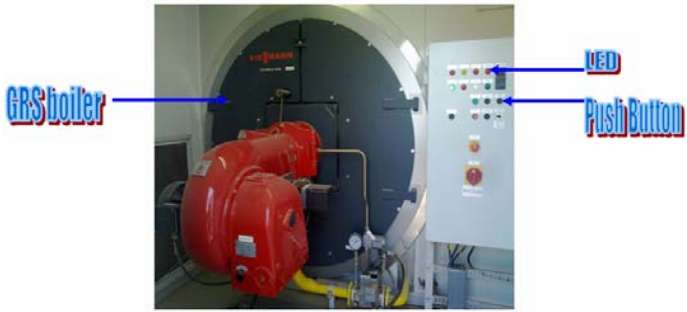

Figure 2. Local control panel of GRS

The ON-OFF switch classified into of two types the push buttons switch and mode selector switch.

Fig. 3 describes the flowchart (sequence of events) of GRS operation. Thus initially the system set to its initial parameters.

Figure 3. Local control panel of GRS

Then, the GRS system enters the stage of energizing the main power by pushing the rest push button. Now the system boiler is ready to start (if there is no faults or an errors), if the system not ready to start the check and reset actions should be repeated again. The boiler of GRS system starts its action (heating Gas), until some desired temperature which has been selected as a set point is achieved. The range of selection of temperature set point is bounded between (0-100) Cº centigrade degree. According to the set point value, the system starts to increase or decrease its temperature, by comparing the desired set point (selected previously) with the actual current temperature. If the actual temperature of GRS boiler less than the set point temperature the control circuit increase the temperature to reach the desired temperature until the actual temperature reaches the set point temperature, at this time the controller waits till temperature drop. When the actual temperature of GRS boiler is very high (high-high) temperature, the controller stops immediately the increment of temperature and shut downing the GRS boiler to prevent the hazards.

The maintenance phase at this level is activated to handle the faults, and then reset the system to enable it to restart again. In the case where the actual temperature of GRS boiler is less than the set point temperature, the system faults must be handled manually by human (system resting) to enable the boiler of GRS to start again. Else the sequence of process waits for temperature dropping.

### C. PIC16F877A Microcontroller Units

PIC-microcontroller is commonly used in robotics and control applications. Its potential use is in data acquisition and measurements [2]. PIC is A Programmable Interface Controller which has wide use areas and are preferred mainly due to low cost, wide availability, free development environments, and easy to access experiences[3]. Fig. 4 shows the pin schematics of (PIC16F877A) produced by Microchip Technology Inc. It has a features like, high performance RISC (Reduced Instruction Set CPU) architecture, operating frequency is 20MHz, only 35 single word instructions, two 8-bit and one 16-bit Timers, fifteen Interrupts, built in support for SPI, USART, A/D etc.[2], 8K programmable memory & 368 data bytes, five I/O Ports as A, B, C, D, E etc, that is used in this work. [4].The name PIC initially referred to "Programmable interface controller", but shortly thereafter was renamed as Programmable Intelligent Computer PIC (reprogramming with flash memory) capability [5]. This is because the PIC memory is divided into two main parts which is the data memory and program memory. Different bus lines are used to enable both of this memory to communicate with one another. Apart from the advantage of the Harvard architecture is that both of the busses can be used simultaneously in the same clock cycle.
In this architecture memory management is divided into two as follows [3]:

1) Memory data
2) Program memory or Flash Program Memory

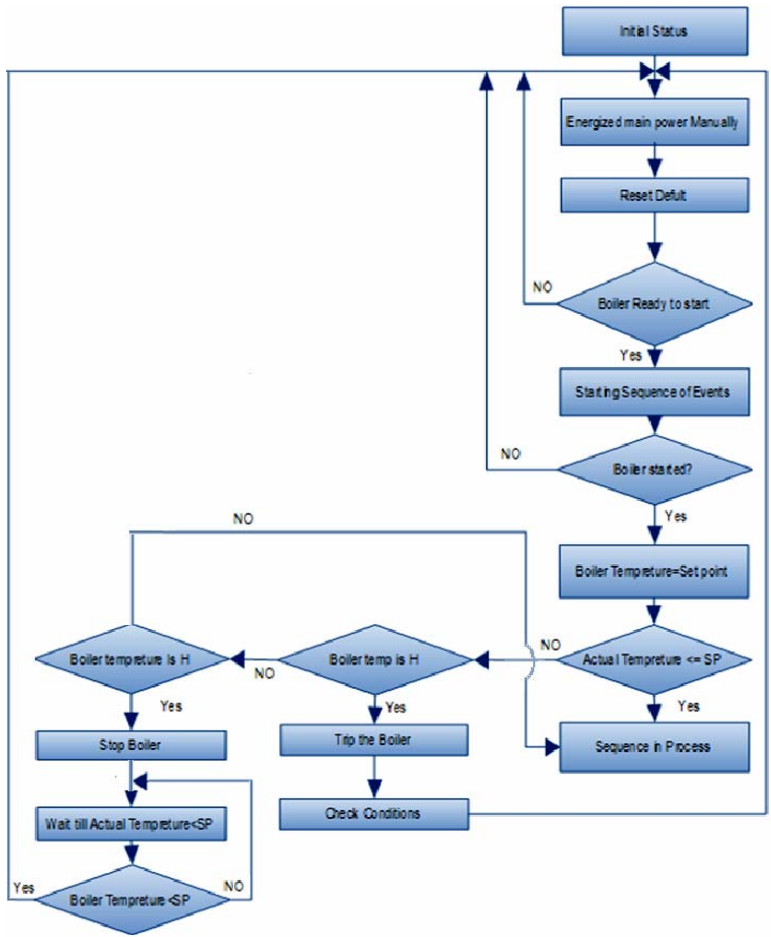

Fig. 4 shows the block diagram for the internal architecture of PIC16F877A microcontroller; there are three main blocks in the internal architecture of the microcontroller [5].

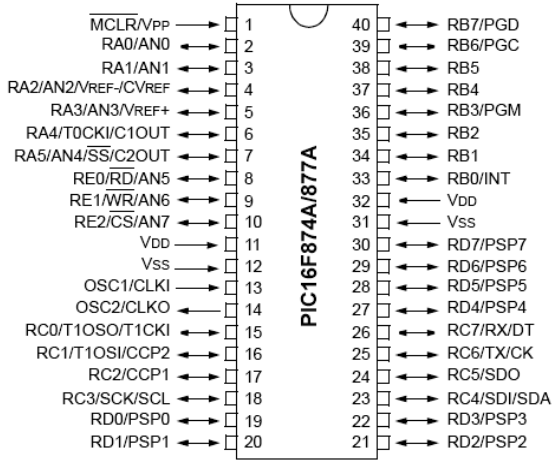

Figure 4. Local control panel of GRS

Those three main blocks are:
1) Flash Program Memory
2) Random Access Memory
3) Electrical Erasable Programmable Read Only Memory

The first part in the internal architecture of the (PIC16F877A) microcontroller is the Flash program memory. The Flash program memory is used to store programs. The advantage is that it can be reprogrammed up to a thousand times. The program that is stored in here will not be erased when the power is cut off.

The second part of the internal architecture of the microcontroller is the file register for the RAM (Random Access Memory). The size of this register is 368 bytes, the data inside it will be erased once the power supply is cut off but it has no limitation to the number of times it can be rewritten [3]. This memory is divided into two sections:

1) Special Function Registers (SFR)
2) General Purpose Register (GPR)

The third section of the internal architecture of this microcontroller is the EEPROM. This microcontroller has 128 or 256 bytes of data EEPROM with an address range from $00_h$ to $FF_h$. The program that has been written into it will not be erased when the power supply is cut off and it can be erased and rewritten up to a million times [3].

The memory data for EEPORM can be written and read during normal operations which are in the range where VDD is full. This memory is not directly mapped into the register files but indirectly addressed to the SFR (Special Function Register) [3]. Four types of SFR used for reading and writing memories are as follows:

1) EECON1
2) EECON2
3) EEDATA
4) EEADR

EECON 1 is the control register for the EEPROM while EECON2 is the protection writing register for the EEPROM. EEDATA proves the space to store data that has the size of 8 bits. EEADR meanwhile is the address marker for the EEPROM. EEPROM data memory allows read and write bytes. If the written bytes are activated then it will automatically erase the old location and thus installs new data. This technique is known as the erase first old data before entering the new data. The writing period is controlled by an on chip timer and it changes according to the voltage, temperature and differs from chip to chip. For components that have a classified code, the CPU can still continue the process of reading and writing in its EEPROM memory [3].

## II. RELATED WORK

Mohd Suhaimi B. Sulaiman and others [2010] design system of GUI Based Remote On/Off control and monitoring Single Phase Lamp (small project comparing with this proposed system) Using Microcontroller [6]. The design of Mohd and others have multiple drawbacks/limitations the important of them the GUI without feedback, more instruction so it take more time, there isn't input signals and no automation,

The motivation reach to design and implement a complete system, i.e. overcome the drawbacks of human (manual tasks) in GRS at EPS and the limitations at Mohd system to achieve a first embedded automation configuration (hardware and software) system that can be used to control all operation/tasks related to GRS at Erbil Power Stations automatically through a GUI and from remote location (about 150m) by using programmable interface controller (PIC16F877A).

## III. THE PROPOSED SYSTEM

The block diagram of proposed system is shown in Fig. 5. It's clear that the signals to/from GRS are exchanged with the PIC through I/O interfacing circuits .Thus the PIC executes the instructions come from GUI and generate control signals to control the proposed machine.

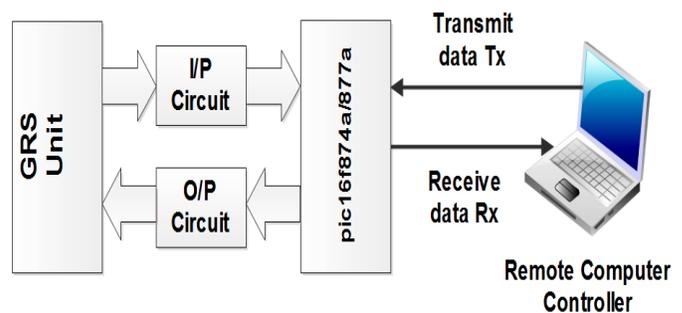

Figure 5. The block diagram of proposed system

The steps followed in designing generic integrated H/W and S/W system are shown in flowchart depicted in Fig. 6. The system design and implementation start by determining H/W and S/W components specifications for the proposed system .After collecting the design related information, the next step involves hardware design in which the hardware components must be compatible with all action done by real GRS system. When the stage of hardware design is complete a review for the hardware design elements is started. If the hardware elements selected are compatible with GRS machines control signals, the software construction will be started if not, the hardware elements selection is repeated till it satisfies the system design requirements. The software presented in this work can be classified into two software techniques:

the GRS machine hardware components are selected and specified for the design and implementation of the proposal automatic system GRS machine. Fig. 7 shows the original local control Panel of GRS that will be converted into full computerized operation and monitoring from remote location.

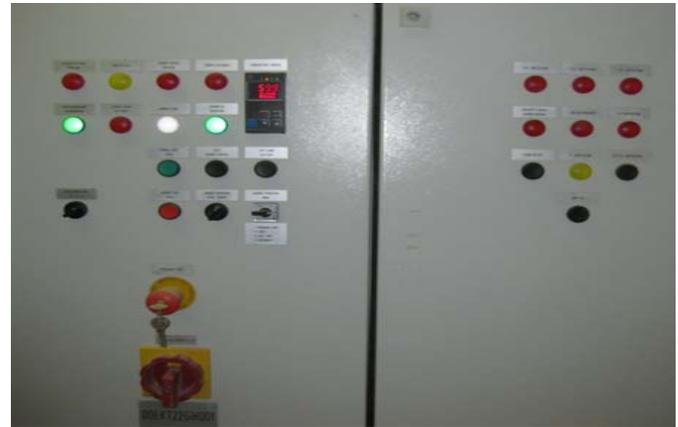

Figure 7. GRS machine control panel.

### B. Collect Design Related Information

In, this step of system design process, the information related to the system operation is collected. The operation environment and the system reaction for the alarm signals and faults represent the main and the first information must be collected. There are two sources of information to be considered about GRS machine operation; the first is the set of documents (manuals) from the GRS manufactory company that contains a list of all possible fault and failure events happened during system operation , the second source of information comes from monitoring of real GRS machine action. Fig. 8 shows the GRS control panel graph in which each component is assigned a number.

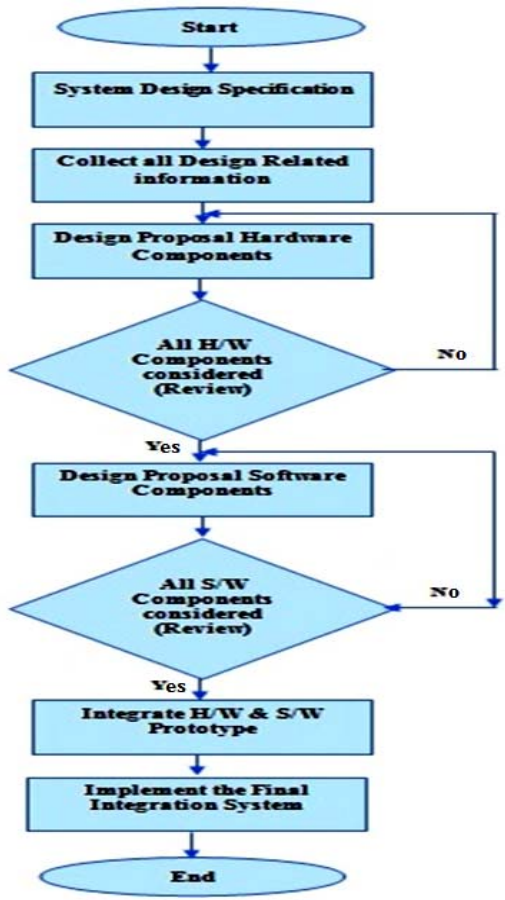

Figure 6. Flowchart of designing generic integrated H/W and S/W design

1. The embedded system software (PIC software) and
2. The interfacing user (GUI) software.

Both of software methods must be test and reviewed. The final step in proposed system design concerned with integrating both of hardware components and software tools. This step will be discussed in details in the next sections:

### A. System Design Specification

The system design specification is the first and most important step in the system design process, in this step

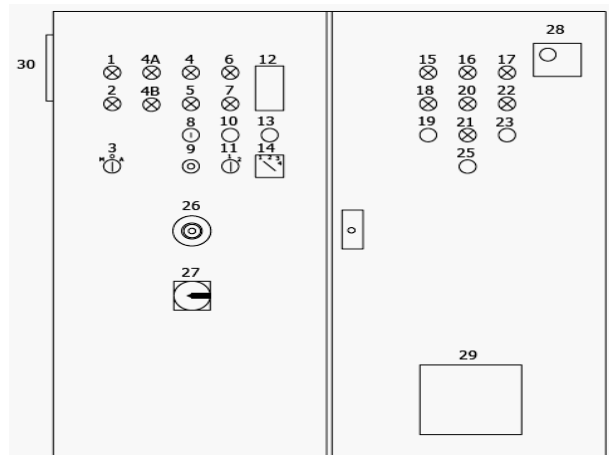

Figure 8. GRS diagram with I/O Signals description

The control signals which handled by the control panel are depicted in Table 1, with their description.

TABLE I.        THE GRS MACHINE SIGNALS

| No.# | (PIC) Pins | Label | Indication | Type | Description |
|---|---|---|---|---|---|
| 1 | RA0 | LED1 | Fault | Digital in | CIRCULATION PUMP OVERLOAD |
| 2 | RA1 | LED2 | Run OK | Digital in | CIRCULATION PUMP IN OPERATION |
| 3 | RA2 | LED4A | Run | Digital in | IGNITION GAS |
| 4 | RA3 | LED4B | Fault | Digital in | LEAKAGE ALARM GAS VALVE |
| 5 | RA4 | LED4 | Fault | Digital in | BURNER MOTOR OVERLOAD |
| 6 | RA5 | LED5 | Run | Digital in | BURNER START |
| 7 | RB0 | LED6 | Fault | Digital in | BURNER DISTURB |
| 8 | RB1 | LED7 | Run | Digital in | BURNER IN OPERATION |
| 9 | RB2 | LED15 | Fault | Digital in | LSA- 00EKT21CL081 |
| 10 | RB3 | LED16 | Fault | Digital in | PSA- 00EKT21CP083 |
| 11 | RB4 | LED17 | Fault | Digital in | PSA+ 00EKT21CP082 |
| 12 | RB5 | LED18 | Fault | Digital in | SAFETY CIRCUIT BURNER CONTROL |
| 13 | RB6 | LED20 | Fault | Digital in | LOW GAS PRESSURE |
| 14 | RB7 | LED21 | Fault | Digital in | TS+ 00EKT21CT081 |
| 15 | RE0 | LED22 | Fault | Digital in | TA+ 00EKT21CT082 |
| 16 | RC0 | SWITCH3 | SELECTOR | Digital out | SELECTOR SWITCH LOCAL/REMOTE |
| 17 | RC1 | | | | |
| 18 | RC2 | Button8 | start | Digital out | BURNER START LOCAL |
| 19 | RC3 | Button9 | stop | Digital out | BURNER STOP LOCAL |
| 20 | RC4 | RESET Button10 | RESET | Digital out | RESET BURNER CONTROL |
| 21 | RC5 | SWITCH11 | SELECTOR | Digital out | BURNER OPERATION LOCAL REMOTE |
| 22 | RC6 | | | | |
| 23 | RC7 | Button13 | TEST | Digital out | TEST FLAME DETECTOR |
| 24 | RD0 | | | | |
| 25 | RD1 | SWITCH14 | SELECTOR | Digital out | BURNER OPERATION MODE |
| 26 | RD2 | | | | |
| 27 | RD3 | Button19 | RESET | Digital out | ALARM RECEIPT |
| 28 | RD4 | Button23 | TEST | Digital out | TEST TA+ 00EKT21CT082 |
| 29 | RD5 | Button25 | TEST | Digital out | LAMP TEST |
| 30 | RD6 | SWITCH26 | EMERGENCY STOP | Digital out | EMERGENCY STOP |

Then a mapping process from GRS Input and Output control signals to PIC Pins is begun. The input and output signals to /from GRS are digital signals (0-1) described as (ON-OFF) states. Each input signal from GRS machine to the PIC microcontroller and each output signal from PIC to the GRS machine will be mapped in to special pin of the PIC. Table 2 shows the PIC pins assignment with corresponding control signals description.

TABLE II.        THE PIC PIN INDENTIFICATION

| Signal NO.# | Signal Description |
|---|---|
| 1 | H11 CIRCULATION PUMP OVERLOAD |
| 2 | H5 CIRCULATION PUMP IN OPERATION |
| 3 | CIRCULATION PUMP SELECTOR SWITCH |
| 4A | H1 IGNITION GAS |
| 4B | H15 LEAKAGE ALARM GAS VALVE AA005 |
| 4 | H12 BURNER MOTOR OVERLOAD |
| 5 | H3 BURNER START |
| 6 | I3 BURNER DISTURB |
| 7 | H4 BURNER IN OPERATION |
| 8 | S2 BURNER START LOCAL |
| 9 | S3 BURNER STOP LOCAL |
| 10 | S8 RESET BURNER CONTROL |
| 11 | S9 BURNER OPERATION LOCAL REMOTE |
| 12 | N1 TEMPERATURE CONTROL |
| 13 | S10 TEST FLAME DETECTOR |
| 14 | S11 BURNER OPERATION MODE |
| 15 | H7 LSA- 00EKT21CL081 |
| 16 | H10 PSA- 00EKT21CP083 |
| 17 | H9 PSA+ 00EKT21CP082 |
| 18 | H14 EM SAFETY CIRCUIT BURNER CONTROL |
| 19 | S7 ALARM RECEIPT |
| 20 | H2 LOW GAS PRESSURE |
| 21 | H6 TS+ 00EKT21CT081 |
| 22 | H8 TA+ 00EKT21CT082 |
| 23 | S5 TEST TA+ 00EKT21CT082 |
| 25 | S12 LAMP TEST |
| 26 | S6 EMERGENCY STOP |
| 27 | Q1 MAIN SWITCH |
| 28 | S1 THERMOSTAT (INSIDE DOOR) |
| 29 | M1 SWITCHBOARD FAN + AIR INLET |
| 30 | AIR OUTLET FILTER |

The (PIC16F877A) has five-Ports categories named like (A,B,C,D,E) denoted as (RA,RB,RC,RD,RE).Each port of these five-Ports contains 8-Pins,hence the total number of Pins in (PIC16F877A) is 40-Pins.According to Table 2 .The input signals from GRS to PIC will be mapped into 15-Pins of PIC ports (RA0-RA5, RB0-RB7, pinE0), while the output signals from PIC to the GRS have been mapped to another 15-Pins of PIC ports (RC0-RC7, RD0-RD6), hence the total number of used Pins from (PIC16F877A) in proposed system becomes 30-Pins out of 40-Pins available in used PIC. Fig. 9 shows the digital input signals from GRS system to PIC and the digital output signals from PIC to GRS system according to Table 2.

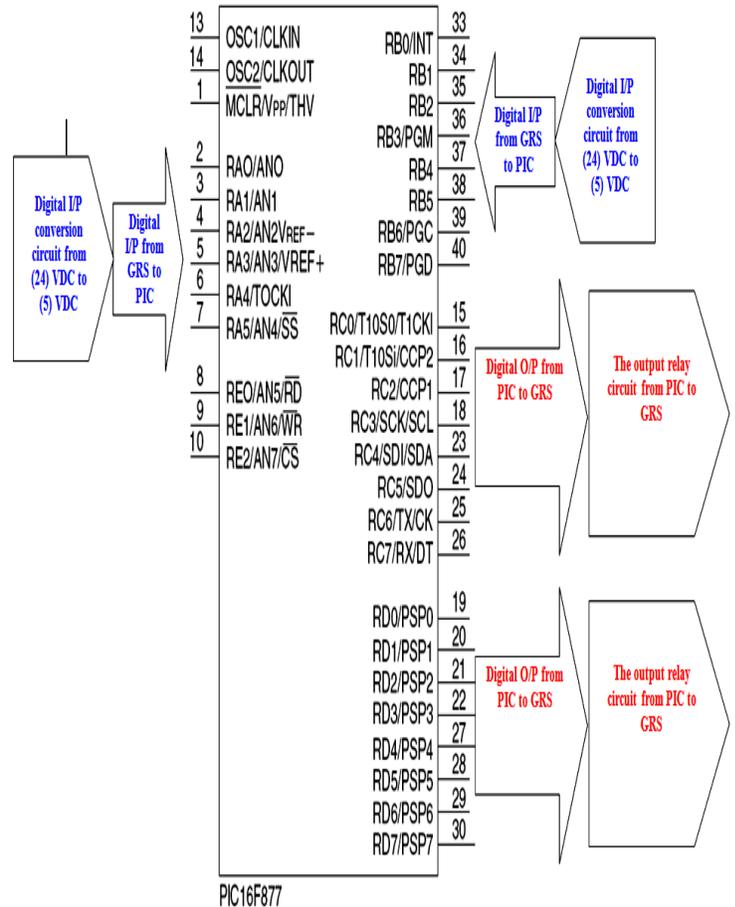

Figure 9. The input and output signals of PIC16F877A

## IV. EXPERMENTAIL RESULTS

Implement the proposed system shown as follow:

### A. The hardware

After collecting all design related information which is necessary for designing a full computerized system that aimed for controlling and monitoring the GRS machine, the next step in the system design implementation is to select the perfect and compatible H/W components to control GRS machine. The development PIC Board KIT EasyPIC6 has been used for programming (PIC16F877A) through USB cable, also the development PIC Board KIT EasyPIC6 will be used as interfacing tool to connect the PIC with the inputs and outputs signals of GRS machine and to connect the PIC with the PC which contains a GUI through (RS232) cable. Input Electronic Interface Circuit from GRS to PIC. The PIC deals with TTL level voltage (0-5) VDC, the (0) VDC represents logic (0) and the (5) VDC represents logic (1) to PIC. Thus the input voltage which has a voltage value more than (6) VDC that might burn the PIC, hence each input signal to PIC from GRS is rated on (24) VDC and this incompatibility in DC voltage level is solved by using Zener diodes (voltage regulator) to protect PIC from high inputs (24) VDC by convert it to (5) VDC .The proposed protection interface H/W is depicted in Fig.10, it contains Zener diodes each of 5.1 V (break down voltage) .A 30KΩ input resistance is connected to input signal to reduce the amount of voltage at the input of each Zener diode into (5) VDC.

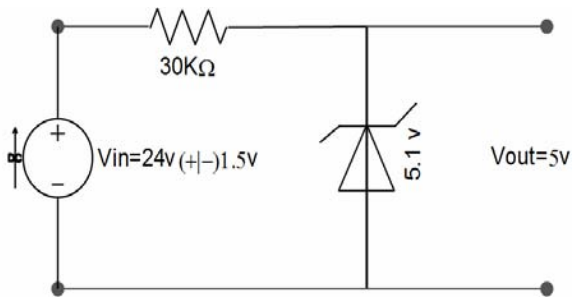

Figure 10. Output voltage limiter circuit

The output relay circuit is connected to provide signals from PIC to GRS. The relay is one of devices used in this proposed design in order to ensure a flexible connection between PIC and the GRS Machine. It's switched ON or OFF according to the level of signals generated from PIC microcontroller. Fig. 11 shows the hardware output circuit design.

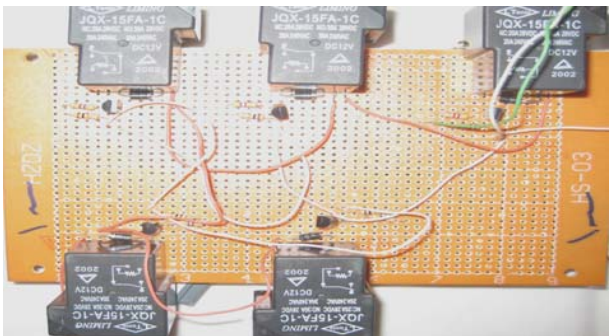

Figure 11. H/W output circuit of relay

The relay circuit replaces the manual switches of GRS with computerized switching, Such that the enable signal from PIC can handle a (230) VAC High Voltage. The outputs enable command for switching ON or switching OFF each relay will be accessed from presented GUI by set or rest push buttons. The relay switch circuit design is depicted as show in Fig. 12.

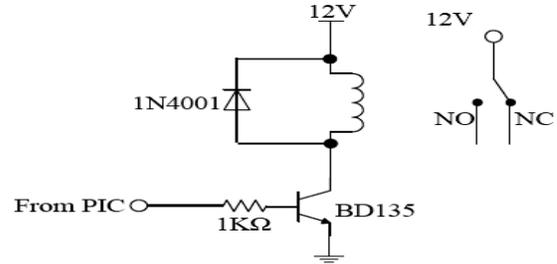

Figure 12. Relay switch circuit

The controller circuit is fabricated on desktop computer case as depicted in Fig. 13.

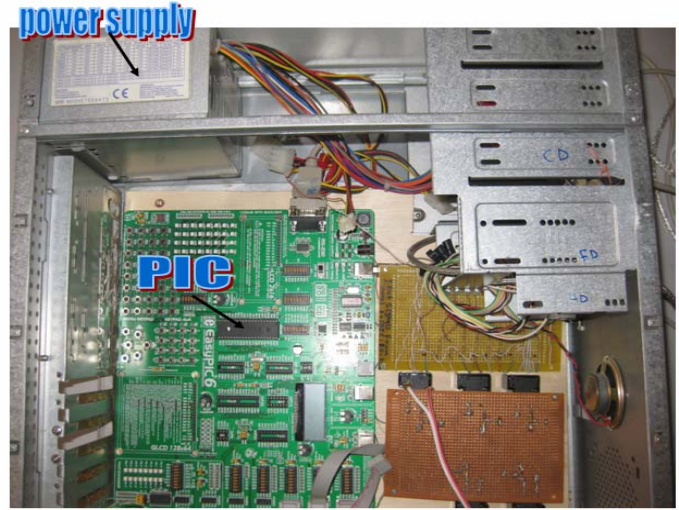

Figure 13. Complete System Fabrication

### B. The software

The embedded programming technique will be invisible programming part to the end user of system which has been written in source code called MPLAB Microchip's Technology Integrated Development Environment based on C-Language codes and environment. The code written has been loaded in the PIC flash memory. The purpose of an embedded system program is to read input data, then processing input through predefined software code, aimed to generate an output signals that control the GRS remotely.

As shown in Fig. 14, the C-Language is a high-level language used for creating the system firmware for low-complexity embedded systems, it is a user-friendly programming technique and it needs only less detailed hardware knowledge. After writing the program in C-Language the MPLAB will edit, check errors and compile the C-language source code into (file. hex) then load it to PIC flash memory. The GUI is implemented using VB, the graphical user interface is a tool that creates an effective communication medium between human and computer; in fact this programming technique of GUI represents the Visible Programming because the user can use the GUI design for remote monitoring and controlling of GRS system.

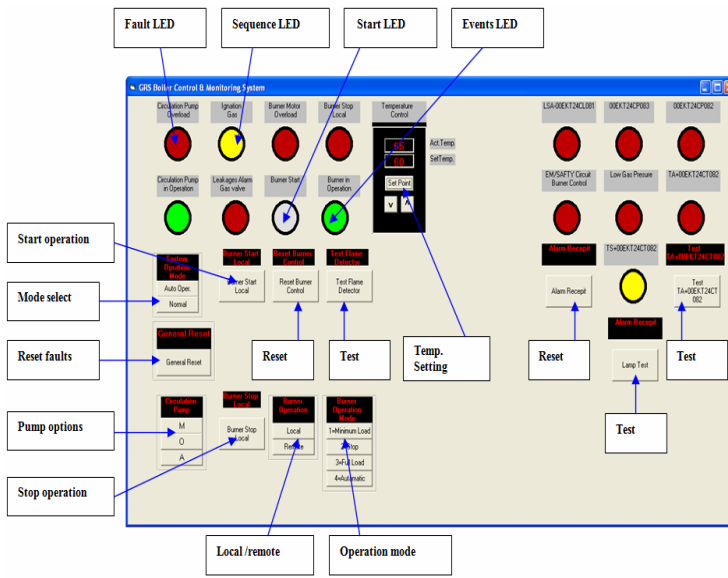

Figure 14. Illustrates C program writing in MPLAB Version 8.33

The GUI design is presented in Fig. 15, the final GUI designed does not need extra training because the user can recognize it easily; this results from the high similarity between it and the original machine.

Figure 15. The GUI design for monitoring and controlling GRS

In, this proposed GUI there are two new push buttons added ,the first is called (Auto operation system mode) which allows a full computerized automatic system mode selection and operation, while the second new push button is called (general reset) which allows resetting all system faults and alarms by just one click. These two functions are not found in the original GRS panel. The proposed GUI is not used only for monitoring it is also used to display the system status through some indicators for input signals and also content a push buttons which acts for output signals. In presented GUI it is possible to use mouse and keyboard for managing and controlling the GRS system. Fig. 15 shows the similarity between the original machine panel and the proposed GUI. C-program is verified and tested by using IDE simulation program.

### C. Remote access through IP network configuration

The aim of using IP network in this paper: *1)* it is open platform (based on TCP/IP) and specifically to satisfy reliability by using TCP [7], *2)* enable GRS remote controlling and monitoring after converting it into full computerize operations. Fig. 16 shows HTML page developed which allows transferring the instructions from GUI toward the GRS system from remote location, about (150m) distance from GRS to control room ,through IP bus network authorized using user name and password.

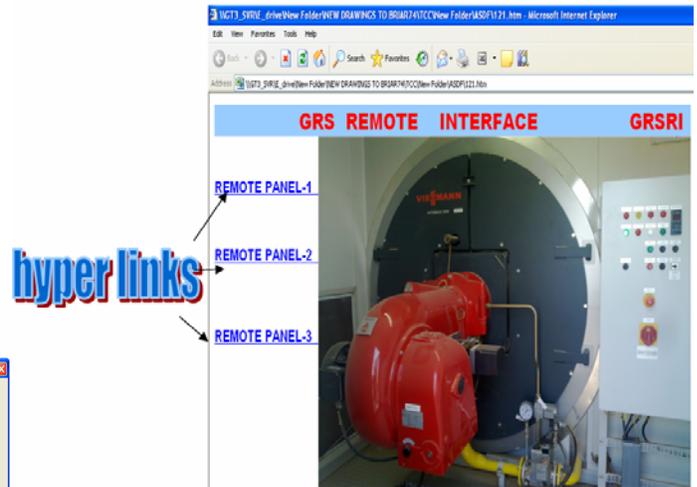

Figure 16. Illustrate HTML page for remote access

Any boiler of GRS can be selected and certain instruction can be transferred to it by clicking on its hyper link text, hence when clicking on remote panel -2 for example the GUI for second GRS is displayed. The HTML page designed to make the access to the remote system easier without needing to the network configuration each time to be accessed by saving the network setting.

### D. Discuss the results

The stage of testing H/W and S/W components of proposed controller is performed before connecting it to the GRS machine. Hence, each system reaction for each predefined event is tested separately, the test stage is important before real system implementation. The S/W level in proposed system is based on Windows platform, performed successfully in transmitting serial data between the computer and the PIC microcontroller. The GUI, based on Windows platform, provides the use of the serial computer port to the system. The GUI has been developed for monitoring and controlling GRS machine from remote location. The GUI uses the command buttons to activate the selected port of the microcontroller that controls the GRS machine as output instruction commands while, the LED indicators refer to the input status as illustrated in Fig. 15. Thus every command button and LED on the GUI represents certain ASCII code from the keyboard. The program waits for another ASCII code to be entered by the mouse.

1) Interfacing GUI with Serial RS-232 and USB Port: The communication port between the PC and the microcontroller can be interfaced either directly via an RS-232 port or the PC USB port. Since computers today are developed with the USB (Universal Serial Bus) port, the GUI based on Windows platforms is designed to be capable of transferring and receiving data via such ports. The USB port of a personal computer is developed to assist the connection of peripheral devices to the computer, improve communication speed and simultaneously support the attachment of multiple devices. The USB-to-RS232 converter is used for interfacing with the USB port of the computer with the system developed. The driver of USB-to-RS232 converter initializes the USB port as a serial port protocol. The use of the converter from a serial interface to the USB port will release a serial communication port to other applications. This allows the devices to be unchanged, making the converter responsible for treating the differences between the protocols. This converter is responsible for transmitting ASCII (American Standard Code for Information Interchange) data from GUI to PIC microcontroller. Each input and output signal is assigned in to different character form ASCII code, the speed rate of bits transmitted and received through the RS232 is 9600 bits per second. The microcontroller compares its reference ASCII code character with the data received and controls the GRS machine when the data received matches the reference ASCII code character which is embedded in the PIC microcontroller. Since data is transmitted using an asynchronous form, the start bit and stop bit indicate the beginning and ending of the data and between the start and stop bit the ASCII code of character in binary form as mentioned before. The input and output signals assignment is shown in Table 3. The signals is mapped into lower case English characters as an example (a, b) have been selected to represent input signals while (y, z) represents output signals.

TABLE III. FOUR CHARACTERS WITH ASCII & BINARY CODE

| char. | Signal | ASCII | Binary |
|-------|--------|-------|---------|
| a | I/O | 97 | 1100001 |
| b | I/O | 98 | 1100010 |
| y | O/P | 121 | 1111001 |
| z | O/P | 122 | 1111010 |

In, the following two figures there are examples of output signal command from PIC microcontroller to GRS through mouse from the GUI. Fig. 17 shows the waveform of ASCII character 'a' which sent by GUI as a "Start" output signal command and the same ASCII character received at the PIC microcontroller port.

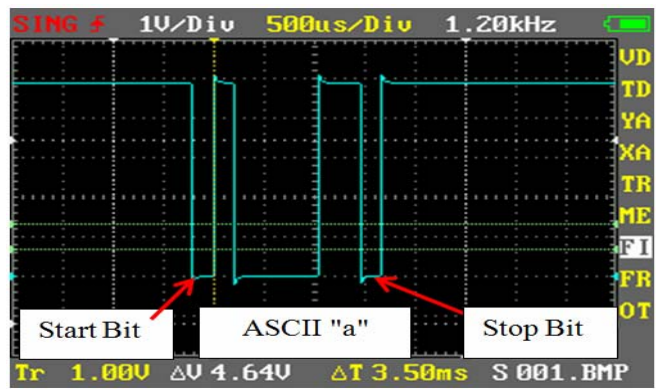

Figure 17. Oscilloscope waveform of character "a"

The output signal command push button which used to "stop" operation of GRS represented by 'b' character Fig. 18 shows the waveform of ASCII character 'b'.

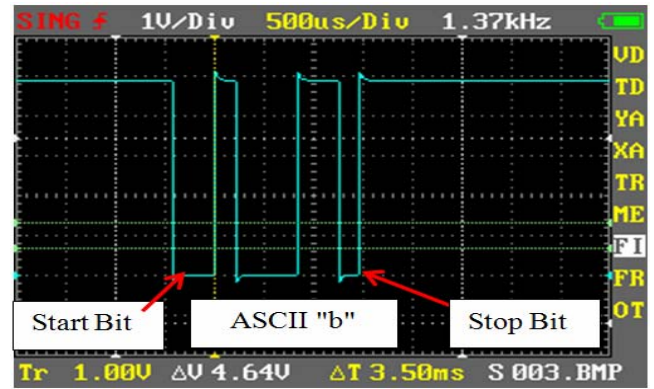

Figure 18. Oscilloscope waveform of character "b"

In the following two figures examples of input signals from GRS machine to PIC microcontroller, a certain LED in the GUI will be set. Fig. 19 shows the waveform of ASCII character 'y' in which a LED that represents "circulation pump over load" is set.

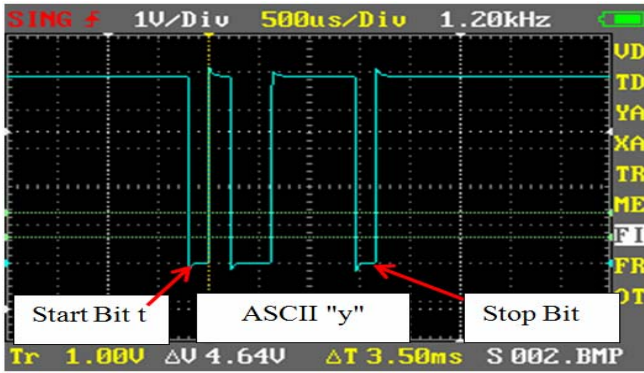

Figure 19. Oscilloscope waveform of character "y"

Fig. 20 shows the waveform of ASCII character 'z' that set a LED on GUI for "circulation pump in operation".

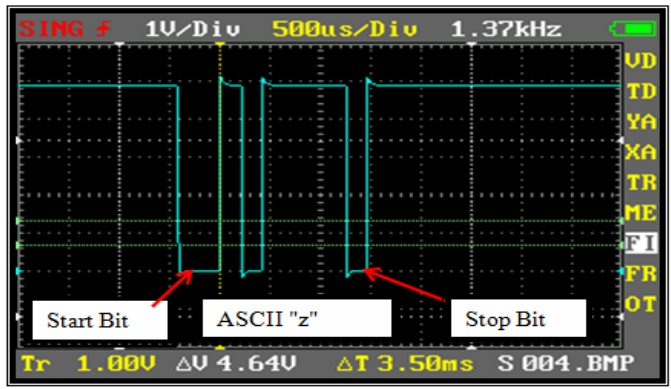

Figure 20. Oscilloscope waveform of character "z"

*2)* Comparison the responses of control local panel and reaction of designed GUI: The proposed system will be tested by making comparison between the responses of control local panel for certain alarm signals and the reaction of the designed GUI. Two scenarios will be considered, the first scenario shows the response of GUI in case of "circulation pump" alarm signal is active ,this will set the green LED in both of GUI and the GRS panel as depicted in Fig. 21 and Fig. 22 which shows the response on the GRS local control panel.

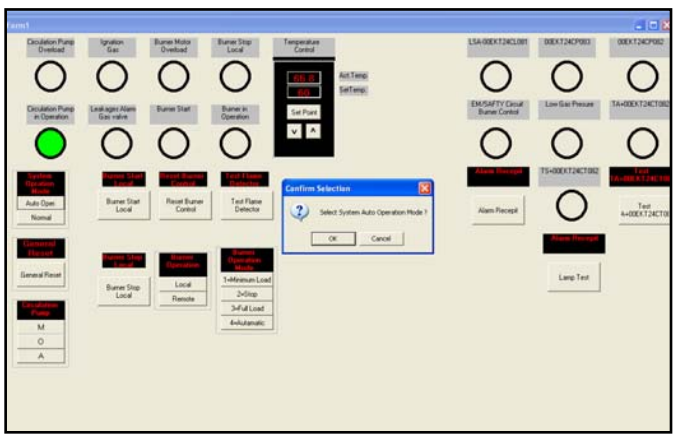

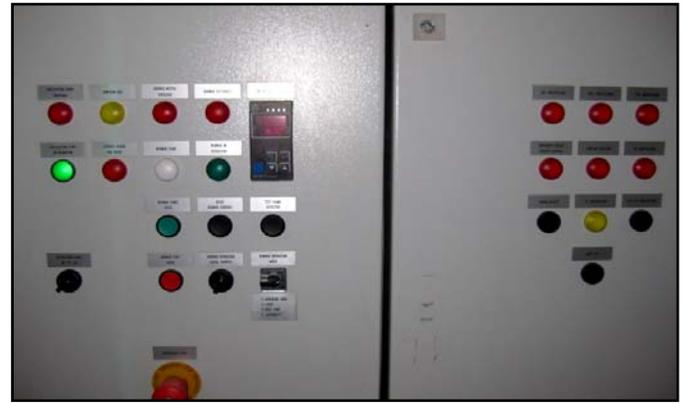

Figure 21. Response of GUI in case of "circulation pump"
Figure. 22. Response of Local Panel in case of "circulation pump"

The second scenario for the GRS machine shows the response of GUI and the local control panel by setting the yellow LED to indicate the activation of "Ignition Gas" alarm signal with white LED setting to indicate the activation of "burner Start "alarm signal . Fig. 23 show the response of GUI, and Fig. 24 shows the response in same

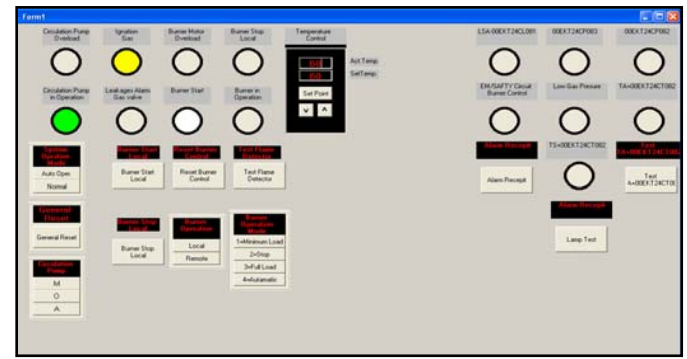

time on the GRS panel.

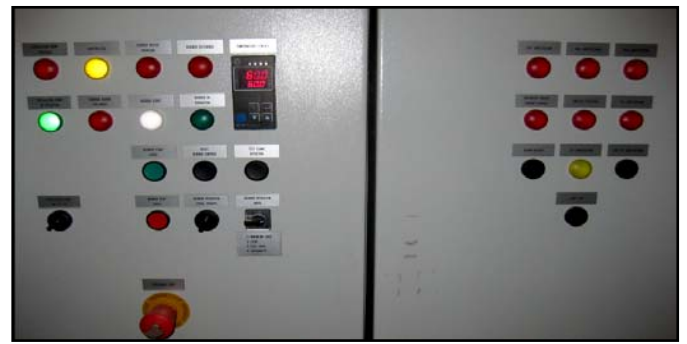

Figure 23. Response of GUI in case of "burner Start"

Figure 24. Response of Local Panel in case of "circulation pump"

The automatic control added in this research is a new function added in the GUI which is not included in the local control panel which enable the operator to control the GRS automatically without human driven control like the auto – pilot system the operator can suspend automatic control and return back to human driven control from remote location . Fig. 25 shows how to set this mode in GUI.

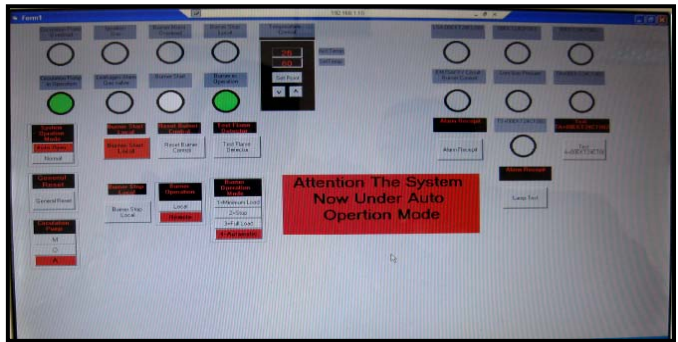

Figure 25. Setting Auto operation mode

A comparison has been made between the proposed system with the nearest related work [6], it found the proposal system in this work outperforms and the GUI of home lighting system by many aspects as shown in Table 4.

TABLE IV.    COMPARATIVE BETWEEN THE PROPOSAL DESIGN SYSTEM OF MOHD AND OTHERS

| Features | The Proposed Design of GRS | System of Mohd and others [6] |
|---|---|---|
| Application | GRS machine in EPS | Single phase lamp |
| GUI Feedback | Include | None, there isn't GUI feedback |
| Remote | Access throughput IP network | Serial cable |
| Embedded S/W | MPLAB version 8.33 based on C | MPASM |
| Instruction Set | Less Instruction | More instruction |
| I/O Signals | Satisfy both I/O Signals | Just output signals |
| GUI Auto Operation | Yes | No |
| GUI Elements | Pushbuttons, indicators shapes & display text box | Only pushbuttons |

## V.    CONCLUSIONS

The functions and performance of the proposed system are tested (realistic) successfully at the EPS. The developed system shows the GUI and the microcontroller is working successfully to control GRS lamps via twisted pair lines. The GUI using VB program is performed excellently in transmitting data (the ASCII character data) to the PIC microcontroller, it can be concluded that GUI using VB can be interfaced with RS232 port of a computer. The operation and monitoring of the GRS machine is huddled and enhanced by utilizing the features of (PIC16F877A) microcontroller, which create a better solution for the GRS problems, so the PIC can be used as an interfacing device between the PC and the machine. The important part in the proposed design is the GUI, the GUI facilitates the engineer work in order to enable a monitoring and controlling of the GRS machine from remote location, hence it is playing a vital role as a interfacing media between the human and the machine which named in the industrial factories and plat as Human Machine Interface HMI.

## ACKNOWLEDGMENT

Thanks a lot to the Erbil Power Stations, Kurdistan-Iraq, the direct Beneficiary and its' director which allowed us to test and implement of proposed system, and then give a certificate for this achievement.

## AUTHORS PROFILE

**Ayad Ghany Ismaeel** received MSC in computer science from the National Center of  Computers NCC- Institute of  Postgraduate Studies, Baghdad-Iraq, and Ph.D. Computer  science in qualification of computer and IP network from University of Technology, Baghdad- Iraq.

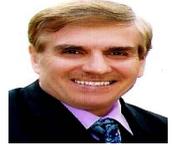

He is professor assistant, at 2003 and currently in department of Information System Engineering in Erbil Technical College- Hawler Polytechnic University (previous FTE Erbil), Iraq. His research interest in mobile, IP networks, Web application, GPS, GIS techniques, distributed systems, bioinformatics and bio-computing. He is senior lecturer in postgraduate of few universities in M.Sc and Ph.D courses in computer science and software engineering as well as supervisor of many M.Sc. student additional Higher Diploma in Software Engineering and computer Science from 2007 till now at Kurdistan-Region, IRAQ. The published papers were in international and national journals about (20) paper.

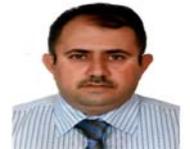

**Raghad Zuhair Yousif** received BSc in Electronics and Communication Engineering from Baghdad University College of Engineering Department of Electronics and Communication Engineering in 1998. Then he received MSc. In Electronics and Communication Engineering from Al-Mustansriyha University in Baghdad College of Engineering Department of Electrical Engineering in 2001. His research was in field of image processing and data security. Then he received a PhD.in Communication Engineering form Department of Electrical and Electronic Engineering from Baghdad University of Technology in 2006. His research was in field of FPGA and channel coding. He had been worked as senior lecturer at Department of Software Engineering College of Engineering Salahaddin University–Hawler from 2006 to 2010.
He is currently Professor Assistant at branch of Communication in Department of Applied Physics College of science at Salahaddin university-Hawler. His research in areas of interest are  Reconfigurable hardware, Channel Coding,  Real Time Systems, Medical Image processing, Remote sensing, Data Security, Bioinformatics and Computer Networks. He is senior lecturer at many MSc Courses for remote sensing, Advanced Computer Networks, Multimedia Technology, Network Security, Real time systems, and supervisor of Many MSc. Thesis in Software Engineering.